\documentclass[preprint,amsmath,amssymb,superscriptaddress,longbibliography,aps,prl]{revtex4-1}
\usepackage{graphicx}% Include figure files
\usepackage{dcolumn}% Align table columns on decimal point
\usepackage{bm}% bold math
\usepackage{xcolor}
\usepackage{color,xcolor}
\usepackage{soul}
\soulregister\cite7 % ??\cite??
\soulregister\citep7 % ??\citep??
\soulregister\citet7 % ??\citet??
\soulregister\ref7 % ??\ref??
\soulregister\pageref7 % ??\pageref??

\begin{document}

\title{Title: Experimental observation of the elastic range scaling in turbulent flow with polymer additives \\
%Short title: Elastic range in turbulence with polymer additives\\
%One sentence summary (Teaser): The elastic range in polymeric fluid turbulence is experimentally observed and its scaling exponent is identified.
}%

\author{Yi-Bao Zhang}%
\affiliation{Institute of Extreme Mechanics and School of Aeronautics, Northwestern Polytechnical University, 710072, Xi'an, China.}

\author{Eberhard Bodenschatz}%
\affiliation{Max-Planck Institute for Dynamics and Self-Organisation, G\"{o}ttingen, D-37077, Germany.}
\affiliation{Institute for the Dynamics of Complex Systems, Georg-August-University G\"{o}ttingen, 37073 G\"{o}ttingen, Germany.}
\affiliation{Laboratory of Atomic and Solid-State Physics and Sibley School of Mechanical and Aerospace Engineering, Cornell University, Ithaca, New York 14853, USA.}

\author{Haitao Xu}%
\email{Email for correspondence: hxu@tsinghua.edu.cn}
\affiliation{Center for Combustion Energy and School of Aerospace Engineering, Tsinghua University, 100084, Beijing, China.}

\author{Heng-Dong Xi}%
\email{Email for correspondence: hengdongxi@nwpu.edu.cn}
\affiliation{Institute of Extreme Mechanics and School of Aeronautics, Northwestern Polytechnical University, 710072, Xi'an, China.}

\date{\today}%

\begin{abstract}

\textbf{
Minute amount of long chain flexible polymer dissolved in a turbulent flow can drastically change flow properties, such as reducing the drag and enhancing mixing. One fundamental riddle is how these polymer additives interact with the eddies of different spatial scales existing in the turbulent flow and in turn alter the turbulence energy transfer. Here we show how turbulent kinetic energy is transferred through deferent scales in the presence of the polymer additives. In particular, we observed experimentally the emerging of a new scaling range, referred to as the elastic range, where increasing amount of energy is transferred by the elasticity of the polymers. In addition, the existence of the elastic range prescribes the scaling of high-order velocity statistics. Our findings have important implications to many turbulence systems such as turbulence in plasmas or superfluids where interaction between turbulent eddies and other nonlinear physical mechanisms are often involved.
}

\end{abstract}

\maketitle
\section{Introduction}\label{sec:intro}
A tiny amount of long chain flexible polymer dissolved in a fluid can drastically change the flow properties.  At low Reynolds numbers, a normal fluid flow is stable and laminar, while the addition of polymers can induce strong fluctuations and create elastic turbulence \cite{Groisman:2000}. In high-Reynolds-number turbulent flows, polymer additives modify the momentum and heat transfer from the wall, resulting in significant reduction of drag \cite{Lumley:1969,Procaccia:2008,White:2008,Samanta:2013} and reduction/enhancement of convective heat transfer\cite{Benzi:2010,Ahlers:2010,Boffetta:2010,Xie:2015,Benzi:2018}. For turbulence in the bulk, far away from the wall, the interaction between polymers and the turbulence energy cascade has long been at the center of research for theoretical reasons, as well as, practical importance \cite{Lumley:1969,Gennes:1986}.

In fully-developed bulk turbulence, kinetic energy is injected into the fluid at the large (forcing) spatial scale $L$, and dissipated by viscosity at the smallest scale in turbulence: the Kolmogorov scale $\eta$ \cite{Kolmogorov:1941}. At intermediate spatial scales $r$ ($\eta \ll  r \ll L$), in the so called inertial range, the kinetic energy is transferred from larger to smaller spatial scales by non-linear interactions. In the inertial range the average  energy flux through scales is constant \cite{Frisch:1995} (Fig. 1A). For turbulent fluid flows with small amounts of polymer additives, it is known that polymers are stretched by the flow and thus draw kinetic energy from turbulence \cite{Perkins:1995}. The loss of kinetic energy is stored as the elastic energy of polymer chains and is either fed back to the flow when polymer-chains recoil  or is dissipated by  polymer-fluid friction or internal interaction within polymer chains. As these physical mechanisms are expected to occur at a wide range of length scales, small amounts of long chain polymers  thus should alter the  energy cascade, i.e. compared to a pure incompressible fluid the inertial range should be cut short and an elastic range should exist (Fig. 1B). Although it was conjectured by de Gennes some thirty years ago that the energy transfer in the small scale portion of the inertial range will be modified by the polymer additives \cite{Gennes:1986}, exactly how the energy is transferred in this modified range is still not known. 

For the turbulence of pure incompressible fluids, such as water, the energy cascade manifests itself in the second order longitudinal velocity structure function (VSF) $S_2(r) \equiv \langle \{[\mathbf{u}(\mathbf{x}+\mathbf{r})-\mathbf{u}(\mathbf{x})]\cdot ( \mathbf{r} / r) \}^2 \rangle$ as $S_2(r)\sim r^{2/3}$ in the inertial range ($\left\langle\ \right\rangle$ denotes spatial and temporal average), as sketched in Fig. 1C, in which also shown is the dissipative range behavior $S_2(r)\sim r^2$ for $r \lesssim \eta$ (as the flow field is smooth at small scales). This notion has been substantiated by extensive numerical and experimental observations (there is a very small correction to the inertial range scaling $r^{2/3}$ due to internal intermittency that we neglect for now) \cite{Frisch:1995}. When polymers are added, as they extract energy from turbulence, they suppress the cascade at some intermediate scale and alter the scaling behavior at smaller scales \cite{Lumley:1969,Gennes:1986,Balkovsky:2001,Fouxon:2003,Ouellette:2009,Kulmatova:2013,Xi:2013}. It is thus natural to expect that as the polymer effects become important, a new scaling range $S_2(r) \sim r^\gamma$ with $2/3 \leq \gamma \leq 2$ would appear, as sketched in Fig. 1D. This new scaling range, however, has not been observed in either experiments or numerical simulations. Numerical simulations inevitably involve modeling simplifications of the polymer-fluid interaction. In addition, they are computationally intensive, even the state-of-the-art simulations show at most a hint of the existence of a new scaling range \cite{Valente:2016}. 
A later theoretical study \cite{Fouxon:2003} predicts that, for scales less than the Lumely scale $r_L$, below which the turbulent fluctuation time scale is faster than the polymer recoiling time, the flow is smooth and the kinetic energy $E(k)$ decays as $k^{-\alpha}$ with $\alpha \geq 3$, which means that the second-order longitudinal VSF follows: $S_2(r) \sim r^\gamma$ with $\gamma = 2$. Recent numerical simulation \cite{Valente:2016}, however, suggests that $ 5/3<\alpha<3$, which is equivalent to $2/3<\gamma<2$. Thus it is crucial to have a more comprehensive measurement of the energy spectra or VSF in turbulent flow with polymer additives to clarify the controversy.
Here we show experimental observation of the new elastic range in a laboratory turbulent flow and the measurement of the scaling of the VSF in the new elastic range, which turns out to be different from any existing theory.

\section{Results}\label{sec:results}
\subsection{The emergence of the elastic range scaling} \label{subsec:elastic_range}
The turbulent flow is generated in a von K{\'a}rm{\'a}n swirling-flow apparatus, which consists of two counter-rotating baffled disks enclosed in a cylindrical tank filled with about 100 liters of water or polymer solutions \cite{LaPorta:2001,Bourgoin:2006,Ouellette:2009,Xi:2013}. The schematic drawing of the system is shown in Fig. S1. The three components of fluid velocity in a central planar region passing through the axis of the tank are measured with a stereoscopic Particle Image Velocimetry (sPIV) system (LaVision GmbH). The measurements show that the flow near the center of the tank is nearly homogeneous and isotropic, for both flows with water and with dilute solutions of long-chain polymers in water. We used polyacrylamide (PAM, with molecular weight $M=18 \times 10^6$ from Polysciences Inc.) in the experiments. The Taylor-microscale-based Reynolds number $R_\lambda=(15u^4/\varepsilon_d\nu)^{1/2}$ for the pure water case is from $340$ to $530$, indicating that there is a fully-developed inertial range in the turbulence. Here, $u$, $\varepsilon_d$, and $\nu$ are the root-mean-square fluctuating velocity, the energy dissipation rate per unit mass by viscosity, and the kinematic viscosity of the fluid, respectively. The parameters of the experiments are summarized in Table S1.
In a solution at equilibrium, the polymers are in the coiled state. If there is flow in the solution but the flow is weak, the polymer will remain in the coiled state by the entropic forces and thus have negligible effect on the flow. If the flow is intense, the polymers will be stretched, and thus store elastic energy and may later release it back into the fluid. The fluid then displays viscoelastic behavior.  
For turbulent flows, this transition is characterized by the Weissenberg number $Wi=\tau_p/\tau_\eta$, which measures the polymer relaxation time $\tau_p$, in terms of the fastest turbulence time scale, $\tau_\eta$. The Weissenberg number must be larger than unity for the polymers to be stretched by the flow. In the experiments reported here, $Wi$ is between $2$ and $11$, which ensures that we observe the viscoelastic effect on turbulence. 
The polymer concentration $\phi$, which varied from $0$ (pure water) to $50$ ppm (parts per million by weight), is below the overlap concentration ($200$ ppm for PAM \cite{Liu:2009}), hence the polymer solution used can be considered dilute, which means that only the interaction between the fluid and single polymer needs to be considered and the direct polymer-polymer interaction can be neglected.
%Here, $u$, $\varepsilon_d$, $\nu$,  $\tau_p$, and $\tau_\eta$ are the root-mean-square velocity, the energy dissipation rate per unit mass, the kinematic viscosity of the fluid, the relaxation time of the polymer and the Kolmogorov time scale, respectively.

To probe the effect of polymers on the turbulent energy cascade, we plot $S_2(r)$ at various polymer concentration $\phi$, shown as symbols in Fig.~\ref{fig:Dll}A. The finite size of our measurement volume allowed us to observe only up to $r/\eta \approx 1000$, which is well within the inertial range and below the forcing scale $L$. Here $S_2(r)$ is averaged over all directions to recover isotropic properties \cite{Hill:2002, Taylor:2003}. For the pure water case ($\phi=0$ ppm), there exists an inertial range scaling $S_2(r) \sim r^{2/3}$ for $r/\eta \gtrsim 200$. When $\phi$ is increased from $\phi = 0$ to $\sim 15$ ppm, $S_2(r)$ is suppressed at the small scale end of the inertial range, which is consistent with previous observations at low concentrations \cite{Ouellette:2009,Xi:2013}. What is new here is that, for higher concentrations $\phi \geq 20$ ppm, a new scaling range $S_2(r) \sim r^\gamma$ with $\gamma = 1.38 \pm 0.04$ appears between the inertial and the dissipative ranges, while the larger scales in the inertial range remain unchanged. The higher the polymer concentration, the more pronounced is the new scaling range, while the scaling exponent $\gamma$ remains the same. We refer to this $r^{1.38}$ scaling range as the elastic range because in this range part of the kinetic energy of the turbulence is diverted into the elastic energy of the polymers \cite{Gennes:1986,Xi:2013}. The emergence of the elastic range becomes clearer in the compensated structure function $S_{2,p}(r) / r^{1.38}$ as shown in Fig.~\ref{fig:Dll}B, where the elastic range appears as a plateau. Here, the subscript $p$ denotes the case of the dilute polymer solution. The plateau, once it becomes visible, broadens with the increase of $\phi$. At the highest concentration $\phi = 50$ ppm, the plateau extends to over about a decade, which convincingly points to the existence of the elastic range. This elastic scaling is also observed in turbulent flows at different Reynolds numbers and Weissenberger numbers,  as well as for different types of polymers (see Figs. S2-S4).

\subsection{Extension of the Batchelor's parameterization} \label{subsec:batchelor}
To quantify the boundary of the elastic range, we adapt an analytical form of $S_2(r)$ for Newtonian turbulence proposed by Batchelor \cite{Batchelor:1951}:
\begin{equation}\label{eq:Dll_n}
  S_{2}(r)=s_{xx}^2 \frac{r^2}{[1+(r/a_1)^2]^{1-\xi/2}},
\end{equation}
where $s_{xx}^2 = \langle (\frac{\partial u}{\partial x})^2 \rangle$ is the mean-square of the derivative of the longitudinal velocity, and $\xi = 2/3$ is to ensure the scaling of $S_2(r) \sim r^{2/3}$ in the inertial range. Equation~\eqref{eq:Dll_n}, although not derived rigorously, provides an accurate description of $S_2(r)$ for both the dissipative and the inertial ranges and has been widely used \cite{Sirovich:1994,Lohse:1995,Lohse:1996}. In this Batchelor's parameterization, $a_1$ is the crossover scale between the dissipative and the inertial ranges: $S_{2}(r) = s_{xx}^2 r^2$ in the dissipative range ($r \ll a_1$) because the velocity field  is differentiable in this range, and $S_{2}(r) = a_1^{4/3} s_{xx}^2 r^{2/3}$ in the inertial range ($r \gg a_1$). For Newtonian turbulence, $S_{2}(r) = C_2 (\varepsilon_t r)^{2/3}$ in the inertial range \cite{Kolmogorov:1941,Frisch:1995}, where $\varepsilon_t$ is the energy transfer rate per unit mass and $C_2$ is the Kolmogorov constant. Clearly $C_2 (\varepsilon_t)^{2/3} = a_1^{4/3} s_{xx}^2$. Therefore the value of $a_1$ can be determined as $a_1 / \eta = (15 C_2)^{3/4} \approx 13$ by the exact relation $s^2_{xx} = \varepsilon_d/(15\nu)$ for isotropic turbulence, the definition of the Kolmogorov scale $\eta = (\nu^3/\varepsilon_d)^{1/4}$, and the fact that $\varepsilon_d = \varepsilon_t$ for fully-developed Newtonian turbulence.

For turbulent flows with polymer additives, we extend Eq.~\eqref{eq:Dll_n} to:
\begin{equation}\label{eq:Dll_p}
S_{2,p}(r)= s_{xx}^2 \frac{r^2}{[1+(r/a_1)^2]^{1-\frac{\gamma}{2}}}\frac{1}{[1+(r/a_2)^2]^{\frac{(\gamma-\xi)}{2}}},
\end{equation}
where $\gamma = 1.38$, $\xi = 2/3$, and $a_1$ and $a_2$ are the dissipative-elastic range and the elastic-inertial range crossover scales with $a_1 \leq a_2$. This extended Batchelor's parameterization (Eq.~\eqref{eq:Dll_p}) is reduced to Eq.~\eqref{eq:Dll_n} when $a_1 = a_2$. Note that $s^2_{xx}$, $a_1$ and $a_2$ are unknown parameters that depend on the polymer concentration $\phi$ and $R_\lambda$. In the inertial range, Eq. ~\eqref{eq:Dll_p} is reduced to 
\begin{equation}\label{eq:Dll_IR}
S_{2,p}(r) = a_1^{2-\gamma} a_2^{\gamma - \xi} s_{xx}^2 r^\xi = a_1^{0.62}a_2^{0.71}s_{xx}^2 r^{2/3}, \quad (r \gg a_2), 
\end{equation}
which should recover the inertial range scaling for Newtonian turbulence $S_{2}(r) = C_2  (\varepsilon_t r)^{2/3}$. This puts a constraint on these parameters: 
\begin{equation}\label{eq:constraint}
a_1^{2-\gamma} a_2^{\gamma - \xi} s_{xx}^2 = C_2 \varepsilon_t^{2/3}. 
\end{equation}

We then fit Eq.~\eqref{eq:Dll_p} with the experimental data at different polymer concentrations. The fits are shown as solid curves in Fig.~\ref{fig:Dll}A, which follow the experimental data well in the entire range and in all cases, suggesting that Eq.~\eqref{eq:Dll_p} captures quantitatively the structure function $S_{2,p}(r)$. The values obtained directly from the fitting show that as $\phi$ increases, the crossover scale $a_1$ decreases while $a_2$ increases, leading to a widening of the elastic range.

The elastic range scaling given by Eq.~\eqref{eq:Dll_p} gives an exact form of $S_{2,p}(r)$ in the elastic range:
\begin{equation}
S_{2,p}(r) = a_1^{2-\gamma} s_{xx}^2 r^\gamma = a_1^{0.62} s_{xx}^2 r^{1.38}, \quad (a_1 \ll r \ll a_2).
\label{eq:S2_elastic}
\end{equation}
Figure~\ref{fig:Dll}C shows the structure function $S_{2,p}(r)$ compensated by Eq.~\eqref{eq:S2_elastic} as a function of $r/a_2$. It is seen that almost all the data sets collapse to a master curve that first shows a plateau in $a_1 \ll r \ll a_2$ then decreases for $r/a_2 \gtrsim 1$. The plateau, which extends almost for one decade, is the elastic range.  The reason that the low concentration  data ($\phi = $ 5 and 10 ppm) do not follow the plateau is that the elastic range is not pronounced in those small $\phi$ cases (as shown in Fig. 2B). 

Note that in the inertial range $S_{2,p}(r)/(a_1^{2-\gamma}s_{xx}^2 r^{\gamma}) = (a_1^{2-\gamma} a_2^{\gamma - \xi} s_{xx}^2 r^\xi)/(a_1^{2-\gamma}s_{xx}^2 r^{\gamma}) = (r/a_2)^{\xi - \gamma} = (r/a_2)^{-0.71}$,  thus all the data should collapse together and follow $(r/a_2)^{-0.71}$ for $r/a_2 \gtrsim 1$. This is indeed the case as shown in Fig.~\ref{fig:Dll}C, where all the data follow the solid curve of $(r/a_2)^{-0.71}$ when $r/a_2 \gtrsim 1$. It also indicates that the scales $r \gtrsim a_2$ are not affected by polymers. This lack of effect was observed in earlier experiments at lower concentrations \cite{Ouellette:2009,Xi:2013}. Here Fig.~\ref{fig:Dll}C clearly shows that $a_2$ captures the transition from the turbulence cascade to the elastic range. As the polymer concentratioon $\phi$ increases, the ratio $a_2/a_1$ increases, which makes the elastic range wider, but the curves at $r \gtrsim a_2$ do not change with $\phi$ any more as long as $a_2 / a_1$ is large enough to form a plateau.

As the polymer concentration increases, the measured $s^2_{xx}$ decreases, which means that the energy dissipated by viscosity at very small scales $\varepsilon_d \sim \nu s_{xx}^2$ decreases with the increase of $\phi$, consistent with earlier experiments \cite{Ouellette:2009,Xi:2013} and numerical simulations \cite{Perlekar:2010}. This decrease of the viscous dissipation $\varepsilon_d$ with $\phi$, together with the independence of the turbulence energy transfer rate $\varepsilon_t$ at larger scales $r \gg a_2$, indicates that the energy transfer rate in the elastic range must vary with scale $r$ for the range $a_1 < r < a_2$ in a non-trivial way, which we will investigate next.

\subsection{Measurement of the kinetic energy transfer rate} \label{subsec:energy_tran_rate}

The energy transfer rate can be obtained from an exact relation in the inertial range known as Kolmogorov's four-fifths law: $S_3 (r) = (-4/5)\varepsilon r$, where $S_3 (r) \equiv \langle \{[\mathbf{u}(\mathbf{x}+\mathbf{r})-\mathbf{u}(\mathbf{x})]\cdot ( \mathbf{r} / r) \}^3 \rangle$ is the third-order longitudinal VSF. In Fig.~\ref{fig:third_order} we plot the local energy transfer rate defined as
\begin{equation}
\varepsilon(r) = -(5/4) S_{3}(r)/ r
\label{eq:epsilon}
\end{equation}
 as a function of $r$ for different polymer concentrations $\phi$. The plateau of the data of $\phi = 0$ ppm clearly shows that the four-fifths law holds for the pure water case in our flow. 
When polymers are present, the local energy transfer rate measured by Eq.~\eqref{eq:epsilon} is not a constant, rather, it varies as $\varepsilon_p(r) \sim (r/a_2)^\beta$ with $\beta = 1.10 \pm 0.07$ in the elastic range $a_1 \ll r \ll a_2$. This dependence of $\varepsilon_p$ on $r$ is consistent with the observed scaling of $S_{2,p}(r)$ in the elastic range: If we adopt a local energy transfer rate for $S_{2,p}(r)$, we obtain $S_{2,p}(r) \sim (\varepsilon_p r)^{2/3}$ and, when compared to the observed $S_{2,p}(r) \sim r^\gamma$ scaling in the elastic range, that gives $\varepsilon_p \sim r^{\frac{3 \gamma}{2} -1} \sim r^{1.07\pm 0.06}$ for $\gamma = 1.38 \pm 0.04$, in good agreement with $\varepsilon_p \sim r^{1.10\pm0.07}$ measured from the scaling of $S_{3,p}(r)$. We stress here that while Eq.~\eqref{eq:Dll_p} is a parameterization introduced for mathematical convenience, the four-fifths law is exact. The good agreement between the two in probing the local turbulence energy flux gives additional support to Eq.~\eqref{eq:Dll_p}.  %\hl{It should be noted that for polymer solution case the power-law amplitude of the four-fifth law may not be -4/5, but it should be still a constant. Therefore, the energy transfer rate $\varepsilon_p(r)$ in the polymer solution case obtained from the four-fifth law may differ from the real value by a numerical factor, but the obtained scaling relationship $\varepsilon_p(r) \sim r^{1.10}$ is not affected.}

It should be noted that for the cases of polymer solutions, the numerical factor of the four-fifths law might not be exactly $-4/5$, but the scaling $\varepsilon_p(r) \sim r^{1.10}$ obtained from Fig.~\ref{fig:third_order} should not be affected. One may wonder whether Eq.~\eqref{eq:epsilon}, which was derived for homogeneous and isotropic turbulence in the Newtonian fluid such as water, is still valid for turbulent flow of polymer solution where the the small scale anisotropy is enhanced with the addition of polymers \cite{Crawford:2008,Xi:2014}. Actually, theoretical study \cite{Hill:2002} and direct numerical simulation \cite{Taylor:2003} have shown that for anisotropic turbulence, isotropic relations can be recovered by spherically averaging the structure functions over all directions for scales below the forcing scale. Thus, Eq.~\eqref{eq:epsilon} can still be used to calculate the average energy transfer rate in the polymer solution case as we used spherically averaged third-order longitudinal VSF. %\hl{One may notice that the inertial scaling range of $S_3 (r)$ (Fig.~\ref{fig:third_order}) starts at a scale smaller than that of $S_2 (r)$ (Fig.~\ref{fig:Dll}) for the pure water case (as shown in Fig. S5 where  $S_2 (r)$ and  $S_3 (r)$ for the same $R_{\lambda}$ are plotted in the same figure), this behavior has been documented in previous measurements \cite{saddoughi1994} and has been theoretically explained \cite{Schumacher2007}.}

Since the energy flux goes to $\varepsilon_t$ when $r \gg a_2$ (inertial range), the local energy flux in the elastic range ($a_1 \ll  r \ll a_2$) should behave as $\varepsilon_p(r) \approx \varepsilon_t (r/a_2)^\beta$ for polymer solution cases, which suggests that less and less energy is transferred down to smaller scales by turbulence non-linear interaction, or equivalently, more and more energy is drawn into the elastic energy of the polymer, the latter is qualitatively consistent with the conjecture by de Gennes \cite{Gennes:1986} and previous experimental results \cite{Xi:2013}. Our results show exactly how this process happens for the case of the interaction between turbulent eddies and polymer elasticity. 

\subsection{The crossover scales $a_1$ and $a_2$} \label{subsec:crossover_scales}

As we have identified the elastic range, it is natural to examine the crossover scale $a_1$ between the elastic range and the dissipation range, and $a_2$ between elastic range and the inertial range and how these two crossover scales vary with the control parameters. Figure \ref{fig:Dll}B suggests that $a_1$ slightly decreases with $\phi$, which, however, is most likely contaminated by the poor spatial resolution of our PIV measurements. The spatial resolution $\Delta x$ of our PIV experiment corresponds to $\Delta x/\eta=5.7$ to 18.7 for the $R_\lambda$ range of $340-530$ (See Table S1). As we have estimated earlier, $a_1/\eta\approx13$ in the pure water case, which is of the same order of our spatial resolution. Thus, $a_1$ obtained from our PIV measurements is inaccurate, as can be seen from the large value of $a_1/\eta \approx 80$ for the pure water case. To correct the errors caused by the low spatial resolution, we have obtained, from the pure water cases with different spatial resolution, a functional form between the discrepancy $\delta_{a_1}$ (between the measured and the theoretical $a_1$) and the relative resolution $\Delta x/\eta$. We then correct the measured $a_1$ by using the above mentioned relationship, the detailed correction procedure can be found in the Supplementary Material. The corrected $a_1$ as a function of $\phi$ for four different $R_\lambda$ is shown in Fig.~\ref{fig:a1a2}, which suggests that dependence of $a_1$ on $\phi$ is very weak for all cases. Also plotted in Fig.~\ref{fig:a1a2} are the data from previous experimental studies using the Lagrangian Particle tracking technique where the spatial resolution is much better (spatial resolution is about $\eta$) and hence no correction on $a_1$ is needed, which shows quantitative agreement with the values of $a_1$ from the PIV data with correction and again supports a very weak dependence of $a_1$ on $\phi$.

As shown in previous subsection, in the elastic range the kinetic energy transfer rate varies as $\varepsilon_p(r) =\varepsilon_t (r/a_2)^\beta$, which means that the energy transfer rate at the crossover scale $r=a_1$ is
\begin{equation}
\varepsilon_p(a_1) =\varepsilon_t (a_1/a_2)^\beta
\end{equation}
which should be the same as the energy dissipation rate in the viscous range $\varepsilon_d =15\nu s_{xx}^2(\phi)$, i.e., 
\begin{equation}
\varepsilon_t (a_1/a_2)^\beta=15\nu s_{xx}^2(\phi)
\end{equation}
This, together with Eq.\eqref{eq:constraint} and the values of $\xi = 2/3$, $\beta = 1.10$, $\gamma=1.38$, yields that 
\begin{equation}
a_1/\eta = (15 C_2)^{\frac{1}{2+ \beta - \gamma}} (a_2/ \eta)^{\frac{\beta+\xi-\gamma}{2+\beta-\gamma}} = (15C_2)^{0.58}(a_2/\eta)^{0.23} ,
\label{eq:a1_with_a2}
\end{equation}
which relates the change of $a_1$ with that of $a_2$. We therefore will return to this after the discussion of $a_2$.

Figure~\ref{fig:a1a2} also shows that $a_2$ increases with $\phi$ with a power-law $a_2\sim \phi^{0.8}$ for the four different $R_\lambda$, and overall $a_2$ is larger for higher $R_\lambda$. Since $a_2$ is the crossover scale between the elastic range and the inertial range, it should be in the same order of the critical length scale $r_\varepsilon$ at which the polymers start to truncate the inertial range \cite{Xi:2013}, i.e., $a_2\sim r_\varepsilon$, with $r_\varepsilon\sim \phi^{0.4}\varepsilon_t^{0.1}\tau_p^{1.1}$ obtained from the balance of the kinetic energy transfer rate per unit mass and the polymer elastic energy transfer rate per unit mass (of the fluid) \cite{Xi:2013}. As the $0.4$ scaling prediction was drawn from experiments with $\phi < 10$ ppm, we checked and found that $a_2$ indeed deviates from 0.8 scaling when $\phi$ is less than $10$ ppm. To compare the behavior of $a_2$ in the low concentration range with the predicted 0.4 scaling relation, we plot a straight line with slope = 0.4 in Fig.~\ref{fig:a1a2}. It is found that our data for $\phi < 10$ ppm indeed shows a hint of $a_2 \sim \phi^{0.4}$. For $a_2 \sim \phi^{0.4}$ and $\phi^{0.8}$, Eq.~\eqref{eq:a1_with_a2} gives $a_1 \sim \phi^{0.09}$ and $a_1 \sim \phi^{0.18}$, respectively. In both cases, the predicted $a_1$ increases with $\phi$. While our data shows roughly a constant value and possibly a very weak decrease of $a_1$ with $\phi$. The reason for this discrepancy between the prediction and the experimental data is not known to us.

%The reason of the different power-law dependence of $a_2$ on $\phi$ ($0.8$ vs. $0.4$) is presumably due to that the latter was based on data in the small concentration range ($0-10$ ppm) and at small Reynolds numbers ($R_\lambda \lesssim 350$), while the former is based on a much larger concentration range ($0-50$ ppm). \hl{Figure \ref{fig:a1a2} shows that indeed $a_2$ continuously decreases in the local gradient with decreasing $\phi$ and the $0.4$ scaling was drawn from the data in this limited range.}

%there appears to be a change of scaling behavior for $a_2$ with $\phi$, especially for lower Reynolds numbers.

It should be noted that Fouxon and Lebedev proposed a dissipation scale $r_\nu = (\nu \tau_p)^{1/2}$, which should be compared with the crossover scale $a_1$ in our study. Note that $r_{\nu}/\eta=(\nu\tau_p)^{1/2}/\eta$ $=(\tau_p/\tau_\eta)^{1/2}=(Wi)^{1/2}$, independent of the polymer concentration $\phi$, which seems to be consistent with our experimental finding that $a_1$ varies very weakly with $\phi$. It should be noted, however, that our measured $a_1/\eta$ for different $R_\lambda$ (hence $Wi$) collapse to each other, which is different from the $r_\nu \sim Wi^{0.5}$ scaling in the Fouxon and Lebedev theory.
 
%\textcolor{red}{We notice that $a_1$ decreases slightly with $\phi$. Overall $a_1$ deceases by 31\% (small $R_\lambda$) to 47\% (large $R_\lambda$). While the variation is much smaller than that of $a_2$, the latter is increased by 628\%. And we believe that this slightly decrease of $a_1$ is due to the poor resolution of our PIV measurements at very small scale. }

It is seen from Fig. 2 that the elastic range can barely be seen when $\phi$ is very small, and it becomes visible and broadens with increasing $\phi$. The reason   is that for small $\phi$, $a_2$ is also very small, thus the scale separation between $a_1$ and $a_2$ is not large enough. While with increasing $\phi$, $a_2$ becomes much larger than $a_1$, thus the elastic range becomes wider and wider. One would expect that an onset of the elastic range should happen when $a_2$ becomes larger than $a_1$. For the pure water case, $a_1 \approx 13 \eta$. Thus if we extrapolate the scaling $a_2(\phi)/\eta \sim \phi^{0.8}$ in Fig. 4 to $a_2/\eta = 13$, we obtain an onset concentration of $0.25$ ppm for $R_\lambda=530$, which is below the smallest concentration used in our experiments.

\subsection{Scaling of the high order velocity structure function}
One key problem of central importance in the study of turbulent flow is the scaling of the high-order VSF in the inertial range $S_{n,w} (r) \sim r^{\xi_w(n)}$ \cite{Frisch:1995}. It is natural to study the scaling of the high-order VSF in the elastic range in turbulent flow with polymer additives $S_{n,p} (r) \sim r^{\xi_p(n)}$. Figure~\ref{fig:high_order}A shows $S_{n,p} (r)$ as functions of $r/a_2$ for $n = 1$ to 8 for the case of $R_\lambda=480, Wi=6.6$ and $\phi=40$ ppm. The elastic range scaling exponents $\xi_p(n)$ of the VSFs are obtained through least square fitting to the data, and are plotted in Fig.~\ref{fig:high_order}C. 
To show that $S_{n,p}(r)$ indeed behaves as power-laws, we compute the local slope of $S_{n,p} (r)$, defined as ${d [\log(S_{n,p}(r))]}/{d [\log(r)]}$, and check if it is a constant in the elastic range. We plot in Fig.~\ref{fig:high_order}B the local slope of  $S_{n,p} (r)$. A flat region (within the two vertical dashed lines) is observed for each order from $n = 1$ to 8, and the averaged value of the flat region of each order is assigned as $\xi_p(n)$. In Fig.~\ref{fig:high_order}C we also plot $\xi_p(n)$ obtained from the local slope method. 
The values of $\xi_p(n)$ measured from the power-law fit and from the local-slope method agree very well. They both increase with $n$, but deviate from a straight line when $n$ is large. The behavior of $\xi_p(n)$ is very similar to that of $\xi_w(n)$, the inertial range scaling exponents of high order VSF for the pure water case, which, as shown in Fig.~\ref{fig:high_order}C, follows the K41 prediction $\xi_w(n) = n/3$ quite well until the order $n$ is large where the intermittency correction becomes significant \cite{Frisch:1995}. 
%\textit{i.e.}, $\xi_w(n) = n/3$ plus intermittency correction, where $n/3$ is the K41 prediction. Also plotted in Fig.~\ref{fig:high_order}C is the inertial range scaling exponents of high order VSF for the pure water case $\xi_w(n)$, it is seen that the data follows prediction of the K41 theory plus intermittency correction. 
This similar behavior implies that, while in the elastic range the energy transfer through scales is altered by polymers, there might still be common features between the Newtonian turbulence and the polymeric turbulence, such as the deviation due to intermittency.

%The difference between the elastic range scaling exponents and the inertial range scaling exponents $\Delta \xi$ presented in the Fig. 3B in the main text are obtained from the data presented in Fig.~\ref{fig:high_order}B.

For Newtonian turbulence, it is realized that intermittency is a manifestation of the strong fluctuation of the instantaneous local quantities \cite{Frisch:1995}. Classical theory \cite{K62} relates the two-point velocity difference with the local energy transfer rate as $\delta_r u \sim (\varepsilon_{t,r} r)^{1/3}$, where $\delta_r u$ is the velocity difference between two point separated over a distance $r$ and $\varepsilon_{t,r}$ is the local energy transfer over the distance $r$. The usual energy transfer rate $\varepsilon_t$ that we refer to can be viewed as $\varepsilon_t = \varepsilon_{t,L}$. The scaling of VSFs are then
\begin{equation}
S_{n,w} (r) = \langle (\delta_r u)^n \rangle \sim \langle (\varepsilon_{t,r} r)^{\frac{n}{3}} \rangle \sim (\varepsilon_t r)^{\frac{n}{3}}(r/L)^{\theta(\frac{n}{3})} \sim r^{\xi_w(n)} ,
\end{equation}
where $(r/L)^{\theta(\frac{n}{3})}$ quantifies the variation of $\varepsilon_{t,r}$ with scale $r$, which is referred to as the intermittency effect \cite{Nelkin:1994}. The scaling exponents are thus $\xi_w(n)= n/3+\theta(n/3)$, of which $\theta(n/3)$ is the intermittency correction.
For turbulence of polymer solutions, similarly we have $\delta_r u \sim (\varepsilon_{p,r} r)^{1/3}$, where $\varepsilon_{p,r}$ is the local energy transfer rate over distance $r$ and $\varepsilon_p = \varepsilon_{p,L} $. Now, if we further assume that the intermittency effect can still be quantified as
\begin{equation}
\langle \varepsilon_{p,r}^n \rangle \sim \varepsilon_p^n (r/L)^{\theta(n)}
\end{equation}
then it follows
\begin{equation}
S_{n,p} (r) = \langle (\delta_r u)^n \rangle \sim \langle (\varepsilon_{p,r} r)^{\frac{n}{3}} \rangle \sim (\varepsilon_p r)^{\frac{n}{3}}(r/L)^{\theta(\frac{n}{3})} \sim r^{\xi_p(n)} .
\end{equation}
Recall that we show already that the energy transfer rate in the elastic range scales as $\varepsilon_p \sim r^\beta$. With this, we obtain an expression for the scaling exponents of the VSFs in the elastic range:
\begin{equation}
\xi_p (n)= (1+\beta)\frac{n}{3}+\theta(\frac{n}{3}),
\end{equation}
which gives a simple relation between $\xi_p (n) $ and $\xi_w (n)$ as
\begin{equation}
\xi_p (n) - \xi_w(n) = \beta \frac{n}{3} .
\end{equation}
Figure \ref{fig:high_order}D shows $\Delta \xi (n) = \xi_p(n)-\xi_w(n)$ as a function of $n$ for $R_\lambda = 480$ and $\phi = 40$ ppm. The data indeed follows the straight line of $\Delta \xi (n) =  1.1n/3$, consistent with our earlier measurement of $\beta = 1.10$. The excellent agreement between the data and the prediction again convincingly shows that in the elastic range, although the energy transfer rate is significantly altered by polymers, the fluctuation of the local energy transfer follow the same statistical description as that of the Newtonian turbulence.

\section{Discussion}\label{sec:discussion}

In a previous theoretical study \cite{Fouxon:2003}, Fouxon and Lebedev predicted the existence of an elastic wave scaling range between the new dissipation scale $r_\nu=(\nu\tau_p)^{1/2}$ and the Lumley scale $r_L=(\varepsilon_d\tau_p^3)^{1/2}$ given by the ``time criterion", with $r_L<L$. They predicted that the velocity spectra $E(k)$ obeys a power-law $E(k)\sim k^{-\alpha}$ with $\alpha\geq3$ in the elastic wave range ($r_\nu<r<r_L$). Although the Fouxon and Lebedev theory predicted some power-law scaling in the elastic range, the assumptions and quantitative prediction of the theory are not supported by our experimental observations. Assumption (1) of the theory is that the inertial energy cascade is terminated below $r_L$, while our measured energy transfer rate scales as $\varepsilon_p\sim r^{1.1}$ in the elastic range, indicating that the inertial energy cascade is not completely terminated but only partially suppressed by polymers and the suppression effect is enhanced with decreasing scales. Assumption (2) of the theory is that the flow field is smooth at scales below $r_L$, which implies that the $n$th-order VSF should scale as $r^n$ in the elastic range, while our experiments show that in the elastic range, the second- and third-order VSF scales as $r^{1.38}$ and $r^{2.1}$, respectively, clearly not a smooth field. Assumption (3) of the theory is that polymer chain is mildly stretched by the turbulent flow, while the Weissenberg number in our experiment is $2.4\leq Wi\leq 11$ (see table S1), the corresponding extension is $0.15<R/R_{max} <0.7$ according to the recent simulation (see Fig. 10(a) of \cite{Watanabe:2010}), which suggests that weakly stretching of polymers is not a necessary condition for the emergence of the new scaling. The key prediction of the theory is that the velocity spectrum $E(k)$ obeys $E(k)\sim k^{-\alpha}$ with $\alpha\geq3$ in the elastic wave range ($r_\nu<r<r_L$), while the second-order VSF measured in our experiment scales as $r^{1.38}$ in the elastic range, which implies that the kinetic energy spectrum scales as $E(k)\sim k^{-2.38}$ in the corresponding range, as shown in Fig. S7.

The Fouxon and Lebedev theory assumes that the upper limit of the elastic range is the Lumley scale $r_L$, which, for a given turbulent flow field, is independent of polymer concentration. This assumption is, however, not consistent with later experimental observations \cite{Ouellette:2009, Xi:2013, Quitry:2016, Sinhuber:2018}. Here our experimental results show that the upper crossover scale of the elastic range $a_2$ varies systematically with polymer concentration (see Fig.~\ref{fig:a1a2}), suggesting that polymer concentration plays an important role in the dynamics of turbulence-polymer interaction.

The physical origins causing the inconsistence between the theory and the experiments might be due to the following two reasons: (1) The theory adopted Lumley's ``time criterion" \cite{Lumley:1969},  which suggests that the polymers start to affect the flow once their relaxation time is larger than the characteristic time scale of the flow, while our results are in favor of the prediction from the energy transfer rate balance model which suggests that the polymers start to affect the flow only when the kinetic energy transfer rate is comparable to the elastic energy transfer rate \cite{Xi:2013}. The latter is more realistic as the polymer concentration entered into the model, and is consistent with experimental observations. While the former has no dependence on polymer concentration. (2) The theory assumes that the dissipation scale is $r_\nu = (\nu \tau_p)^{1/2}$, which implies that the viscous energy dissipation rate is $\varepsilon_d \approx \nu / \tau_p^2$. This assumption does not consider the continuous of energy flux across scales, i.e., from the elastic range to the dissipative range, and predicts that the viscous dissipation rate $\varepsilon_d$ is independent of polymer concentration, both of which are questionable from the physics point of view.

In summary, we have experimentally observed the scaling of the elastic range in the turbulent flow with polymer additives. In addition, with the help of this clear scaling range, we are able to measure the turbulent kinetic energy transfer rate in the presence of polymer additives for the first time from an exact relation --- the 4/5-law of turbulence. It is found that the energy flux through the turbulent flow decreases, while the energy flux through the elastic degree of freedom of polymers increases, with decreasing length scale $r$. And this scale dependent energy transfer rate by the flow successfully explained the scaling of the elastic range. We in addition identify the functional form of the scaling of high order velocity structure function in the elastic range. Our study thus shed new lights to the further theoretical/numerical studies on the interaction between elasticity of polymer additives and turbulent eddies. Similar process could very well be happening in other phenomena involving a cascade and another physical mechanism whose effect is scale dependent, such as electromagnetic interactions in plasmas or Alfven waves in superfluids.

\section{Materials and Methods}
\subsection{The von K{\'a}rm{\'a}n swirling flow system}
The turbulent flow is generated in a cylindrical tank by two counter-roating baffled disks (see Fig. S1 and ref. \cite{voth2002jfm}). The cylindrical tank is vertical mounted, it is $636$ mm in height and $480$ mm in diameter. The disks are open-end cylinders $220$ mm in diameter and $50$ mm deep with eight $5$ mm thick vanes internally mounted on the surface to enhance mixing. The wall of the cylinders is made of Plexiglas and is $10$ mm in thickness. The disks are spaced $416$ mm apart and each is driven by a 1.5 kW computer controlled Servo motor. Six stationary radial vanes are installed between the disk and the container wall  to inhibit large-scale rotation of the flow. The disk rotation rate $f$ can be continuously varied from $0$ to $6$ Hz and we choose $f=0.6, 0.9, 1.2$ and $1.6$ Hz in this study. The top and bottom plates of the tank are made of 50 mm thick aluminum. Two spiral channels are machined into each of the top and the bottom plates. Aluminum covers are attached to the top of the upper plate and the bottom of the lower plate. These covers, together with the upper (lower) surface of the top (bottom) plate, serve as cooling channels of the top and the bottom plates. Cold water is pumped to each of the plates through two inlets and flows out through two outlets via a refrigerated circulator (Polyscience 9702) to regulate the temperature of the two plates. The temperature of the fluid inside the tank is regulated through the top and the bottom plates due to the high thermal conductivity of aluminum and the turbulent mixing inside the tank. During the experiments the temperature of the top and bottom plates thus the fluid inside the tank is kept at $25$ $^\circ$C. The temperature stability of the refrigerated circulator is $0.01^\circ$C.

\subsection{The polymer used in the experiments}
The polymer we used in the experiments is polyacrylamide (PAM, from Polysciences) with molecular weight $M = 18\times10^6$. Polyethylene Oxide (PEO, from Sigma-Aldrich) with $M = 8\times 10^6$ is also used to check the universality of the experimental results. Unless otherwise specified, all the experiments were done with PAM. To prepare the polymer solution with a specified concentration, we first fill the tank with deionized water and then add high concentration polymer solution (stock solution) into the tank, we then turned on the disks with a slow rotating rate of $f=0.4$ Hz to mix the stock solution and the deionized water. The stock solution is prepared as follows: prepare 4 liters of deionized water in a beaker, keep stirring the water with a teflon stir at rotation rate of 300 revolution per minutes (rpm), add polymer powders very slowly into the beaker. Three minutes later, the rotation rate of the stir is set to 250 rpm for 1 hour, then 150 rpm for 2 hours and 100 rpm for 3 hours. Finally, the rotation rate of the stir was set to 80 rpm for another 10 hours. In this way, the polymers are well dispersed and dissolved into the deionized water.  The typical concentration of the stock polymer solution is $500$ ppm.

\subsection{The Velocity measurements}
Particle image velocimetry (PIV, from LaVison GmbH) is employed to measure the velocity in a meridian vertical plane at the center of the tank, where the flow is nearly homogenous and isotropic. There are seven circular glass windows (diameter = 10 cm) with flat surfaces allowing laser illumination and optical access of the cameras for the PIV measurements (shown in Fig. S1). The flow is seeded with hollow glass sphere with nominal diameter $d_0=10$ $\mu$m and density $\rho_0=1.1 \times 10^3$ Kg/m$^3$ and the corresponding relaxation time $t_0={\rho_0d_0^2}/{(18\rho\nu)}\approx6.8\times10^{-6}$ s. The Stokes number $S_t=t_0/\tau_\eta$ is much smaller than 1 in our study (see Table S1), indicating that the flow can be faithfully followed by the particles. Here, $\nu$ is the kinematic viscosity of the deionized water at $25$ $^\circ$C and $\tau_\eta$ is the Kolmogorov time scale. The tracer particles are illuminated by green laser light with wavelength = $532$ nm. Two types of PIV measurements were performed, for the polymer concentration dependence of the second order velocity structure function we used the stereo PIV; while for the third order and higher order velocity structure functions which require large number of experimental data, we used the planer PIV with higher spatial resolution and much longer acquisition time. In the stereoscopic PIV measurements, two cameras separated by $60^\circ$ are arranged on the same side of the laser sheet and the field of view is $90\times80$ mm$^2$ ($800\times800$ pixels on each camera). In planar PIV, one camera is mounted normal to the laser sheet and the field of view is about $60\times60$ mm$^2$ ($1664\times1600$ pixels). In both cases, the camera is operated in dual-frame mode, and the time delay between the laser pulses is adjusted according to the rotation rate of the disk $f$ so that the maximum movement of tracer particle is less than $1/4$ of the final interrogation window. The captured image pairs are processed with a two-pass interrogation procedure. In the first pass, the image pairs are interrogated using $48\times48$ pixels interrogation regions and $50\%$ overlap. In the second pass, the image pairs are interrogated using $24\times24$ pixels interrogation regions and $50\%$ overlap. Spurious vectors are detected by median test and replaced by interpolating neighbor vectors. Each snapshot is composed of $78\times70$ and $137\times132$ vectors for stereoscopic and planar PIV, respectively. The spatial resolution (physical distance between two neighboring vectors) are $1.10$ mm and $0.43$ mm,  respectively. Stereoscopic PIV is employed for disk rotating rate $f=0.6$ ($R_\lambda = 340$), $f=0.9$ Hz ($R_\lambda = 400$), $1.2$ Hz ($R_\lambda = 480$) and $1.6$ Hz ($R_\lambda = 530$)  and  15800 velocity maps were taken for polymer concentration: $\phi = 0$ to $50$ ppm. Planar PIV is employed for disk rotating rate $1.2$ Hz ($R_\lambda = 480$)  and 11 runs with total 50600 velocity maps are taken for polymer concentration: $\phi = 0$ to $40$ ppm. Thus the total number of velocity vectors is $50600\times137\times132$. These correspond to $2.7\times10^9$ to $4.4\times10^{10}$ data points of velocity increments for the smallest $r$ to the largest $r$ in the inertial range, they are statistically enough for these high order moments.

\section{Supplementary materials}
Supplementary material for this article is available at http://advances.sciencemag.org/xxxx/\\
\noindent The elastic range scaling in flow with different $R_\lambda$, $Wi$ and different polymer.\\
Kinetic energy spectrum in turbulent flow with polymer additives.\\
The correction to the crossover scale $a_1$.\\
Convergence of high order moments of velocity increments.\\
Fig. S1. Experimental setup.\\
Fig. S2. The elastic range scaling observed in the second order \textit{transverse} VSF $S_2^T(r)$.\\
Fig. S3. The elastic range scaling observed at different $R_\lambda$ and $Wi$.\\
Fig. S4. The elastic range scaling observed with a different type of polymer.\\
Fig. S5. The third- and the second- order longitudinal velocity structure functions for the pure water case.\\
Fig. S6. Power-law amplitude of the 4/5-law for the pure water case.\\
Fig. S7. Kinetic energy spectrum in turbulent flow with polymer additives.\\
Fig. S8. The correction to the crossover scale $a_1$.\\
Fig. S9. Convergence of high order moments of velocity increments.\\
Table S1. Experimental parameters in this study.

\section{References and Notes}
%\bibliography{SA_polymer.bib}

\section*{Acknowledgements}
We are grateful to K.-Q. Xia for sharing the PAM polymer with us, to D. Lohse,  K.-Q. Xia, Chao Sun and S.-D. Huang for stimulating discussions, and to C. Liu and Z.-Q. Zhang for their contributions to the experimental setup. \\
\textbf{Funding}  This work is supported by  the NNSF of China (11772259, 11988102 and 11472094), the 111 project of China (B17037) and the Fundamental Research Funds for the Central Universities of China (No.3102019PJ002). H.-
D. Xi thanks the support from Alexander von Humboldt Foundation which initiated the work presented in this paper.
\\ \textbf{Author contributions:} H.D.X. designed the research, Y.B.Z. performed the experiments, Y.B.Z, E.B., H.X., and H.D.X. analysed the data, and wrote the paper. 
\\ \textbf{Competing interests:}
The authors declare that they have no competing interests. 
\\ \textbf{Data and materials availability:} All data needed to evaluate the conclusions in the paper are present in
the paper and/or the Supplementary Materials. Additional data related to this paper may
be requested from the authors.

\clearpage

\begin{figure}
  \centering
  \includegraphics[width=0.7\columnwidth]{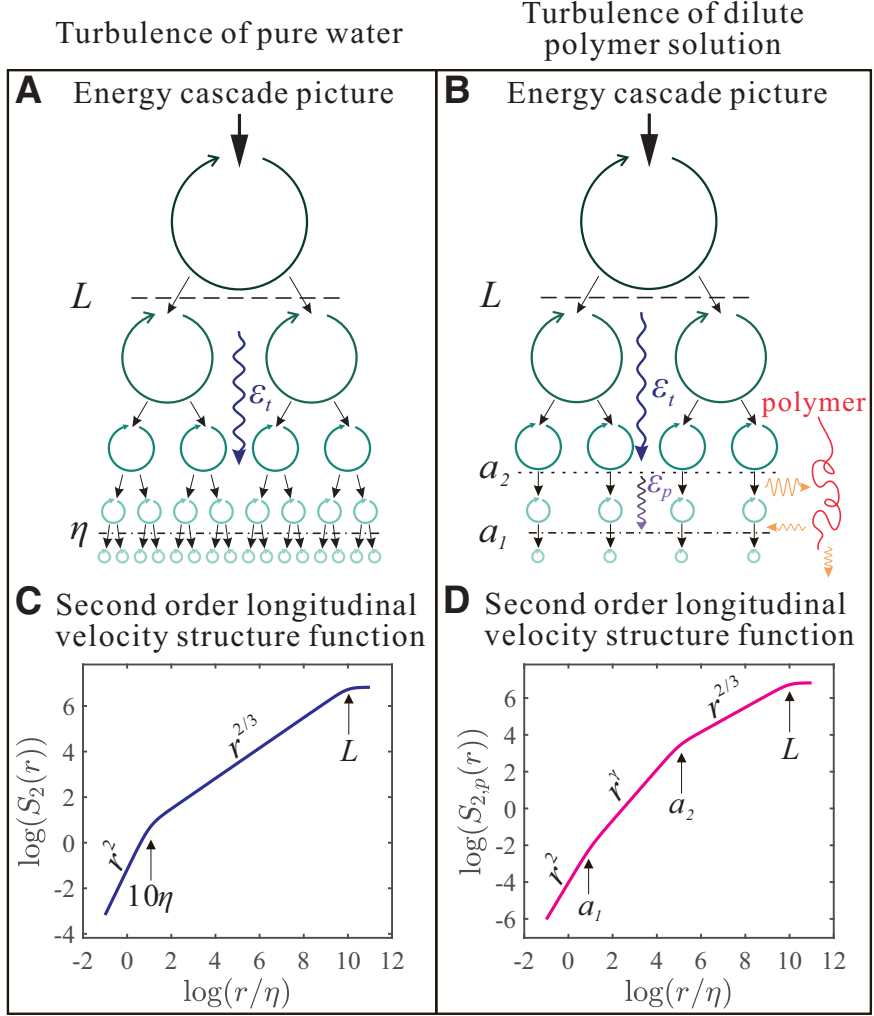}
  \caption{\textbf{Physical picture of the turbulent energy cascade and its manifestation on the second order longitudinal velocity structure function.}  (\textbf{A}) and (\textbf{B}): Cartoons showing the physical picture of the energy cascade in turbulent flow of pure water and dilute polymer solution. (\textbf{C}) and (\textbf{D}): The second order longitudinal velocity structure function in turbulent flow of pure water case and dilute polymer solution case.}\label{fig:cartoon}
\end{figure}

\begin{figure}
  \centering
  \includegraphics[width=0.6\columnwidth]{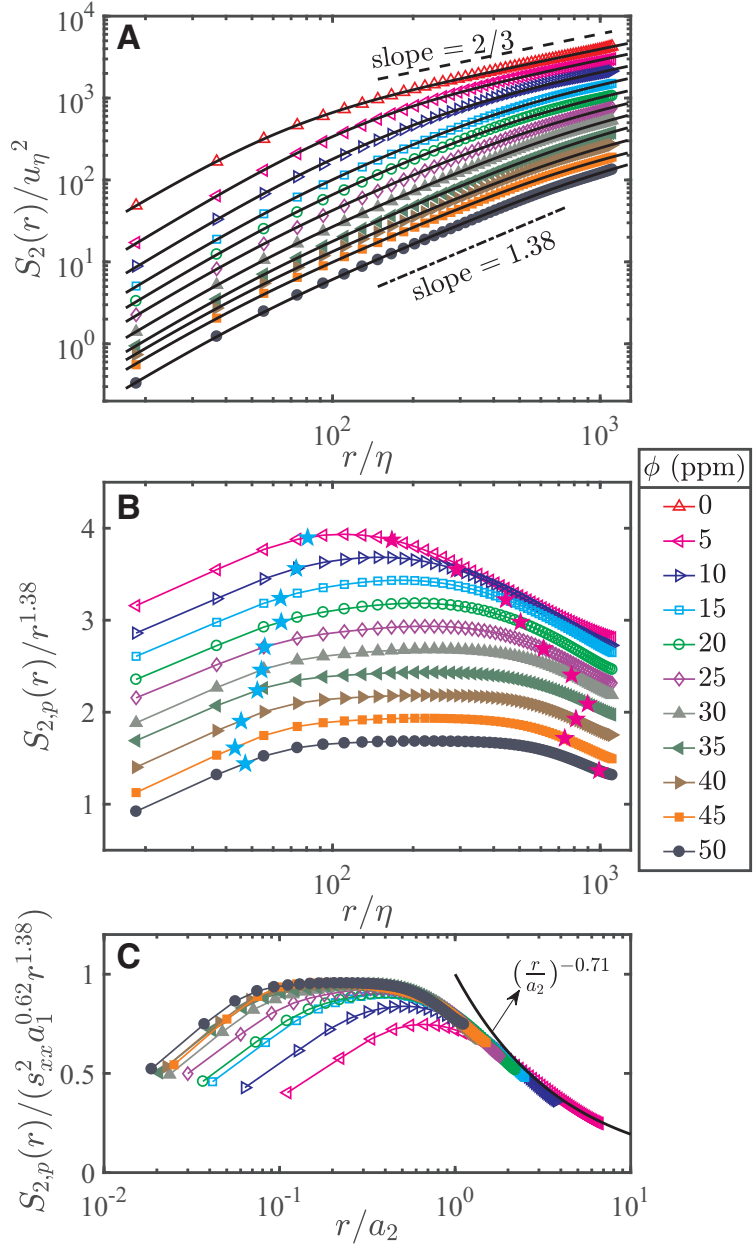}
  \caption{\textbf{Second order longitudinal velocity structure functions $S_2(r)$ for pure water and dilute polymer solutions at $R_\lambda=530$}. (\textbf{A}) $S_2(r)$ and $r$ are normalized by $u_\eta^2$ and $\eta$, respectively. Here, $R_\lambda$, $\eta$ and $u_\eta$ are from the pure water case. The solid curves are fits to the parameterization function (Eq.~\eqref{eq:Dll_p}). For the sake of clarity, lower $\phi$ data has been shifted upwards by $10^{0.15}$ with respect to its higher $\phi$ neighbor. (\textbf{B}) The same data as in (A) but $S_{2,p}(r)$ is compensated by elastic range scaling $r^{1.38}$. For the sake of clarity, each data set has been shifted up by $0.25$ with respect to its higher $\phi$ neighbor. The cyan and magenta pentacles show the crossover scales $a_1$ between the dissipation and elastic ranges, and $a_2$ between the elastic and inertial ranges, respectively. (\textbf{C}) The same data as in (A) but $S_{2,p}(r)$ is compensated by its exact form in the elastic range given by the parameterization: $s^2_{xx}a_1^{0.62}r^{1.38}$,  and $r$ is normalized by $a_2$. The solid curve is $(r/a_2)^{-0.71}$. }\label{fig:Dll}
\end{figure}

\begin{figure}
  \centering
  \includegraphics[width=0.6\columnwidth]{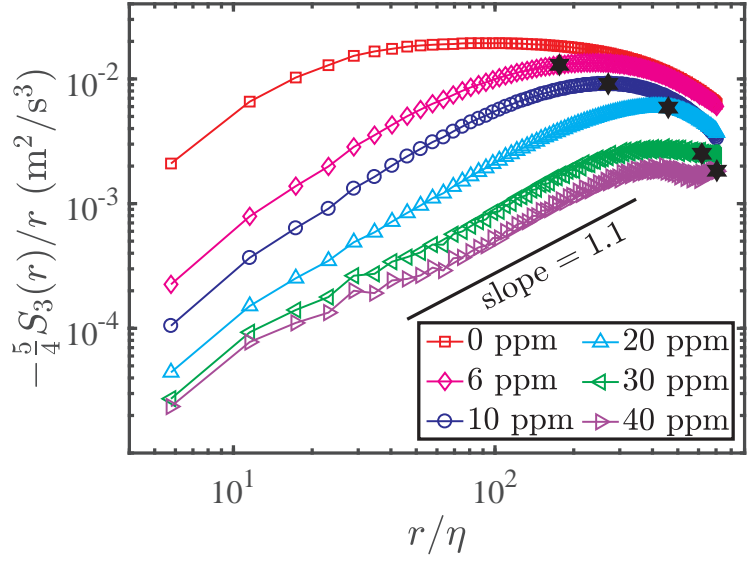}
  \caption{\textbf{The local turbulence kinetic energy transfer rate determined from the third order longitudinal velocity structure function.} Compensated third order longitudinal VSF $-\frac{5}{4}S_3 (r)/r = \varepsilon (r)$ as a function of $r/\eta$ for pure water case and the polymer solution cases at $R_\lambda = 480$. The black pentacles show the crossover scale $a_2$ between the elastic and inertial ranges.}\label{fig:third_order}
\end{figure}

\begin{figure}
	\centering
	\includegraphics[width=0.6\columnwidth]{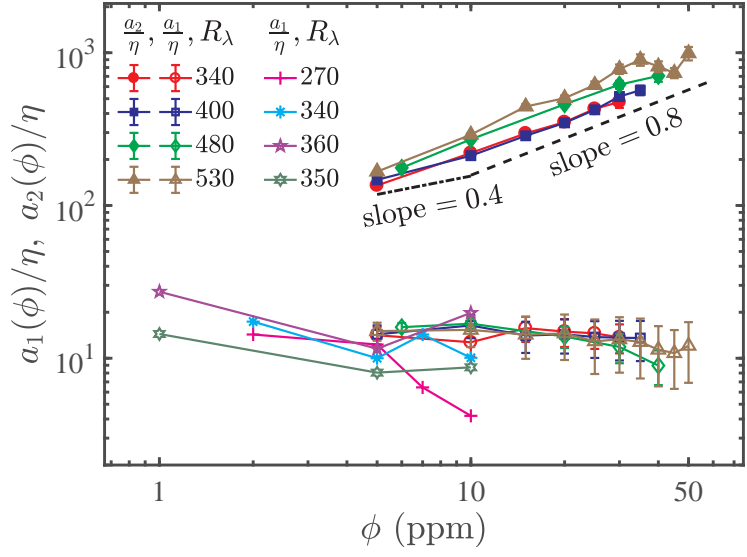}
	\caption{\textbf{The variation of the crossover scales $a_1$ between the dissipation and the elastic ranges and $a_2$ between the elastic and the inertial ranges.} $a_1$ and $a_2$ as functions of $\phi$ for four different $R_\lambda$. Here $a_1$ and $a_2$ are normalized by $\eta$ from the pure water case.  $a_1$ at lower concentration from previous experiments ($R_\lambda = 270, 340, 360$ data from \cite{Xi:2013} and $R_\lambda = 350$ data from \cite{Ouellette:2009}) are also plotted for comparison. The slope = 0.8 straight line is to shown that overall $a_2$ scales with $\phi^{0.8}$, while the slope = 0.4 straight line is to compare the data in the low concentration range with the prediction $r_\varepsilon \sim \phi^{0.4}$ \cite{Xi:2013}.}\label{fig:a1a2}
\end{figure}

\begin{figure}
	\centering
	\includegraphics[width=0.9\columnwidth]{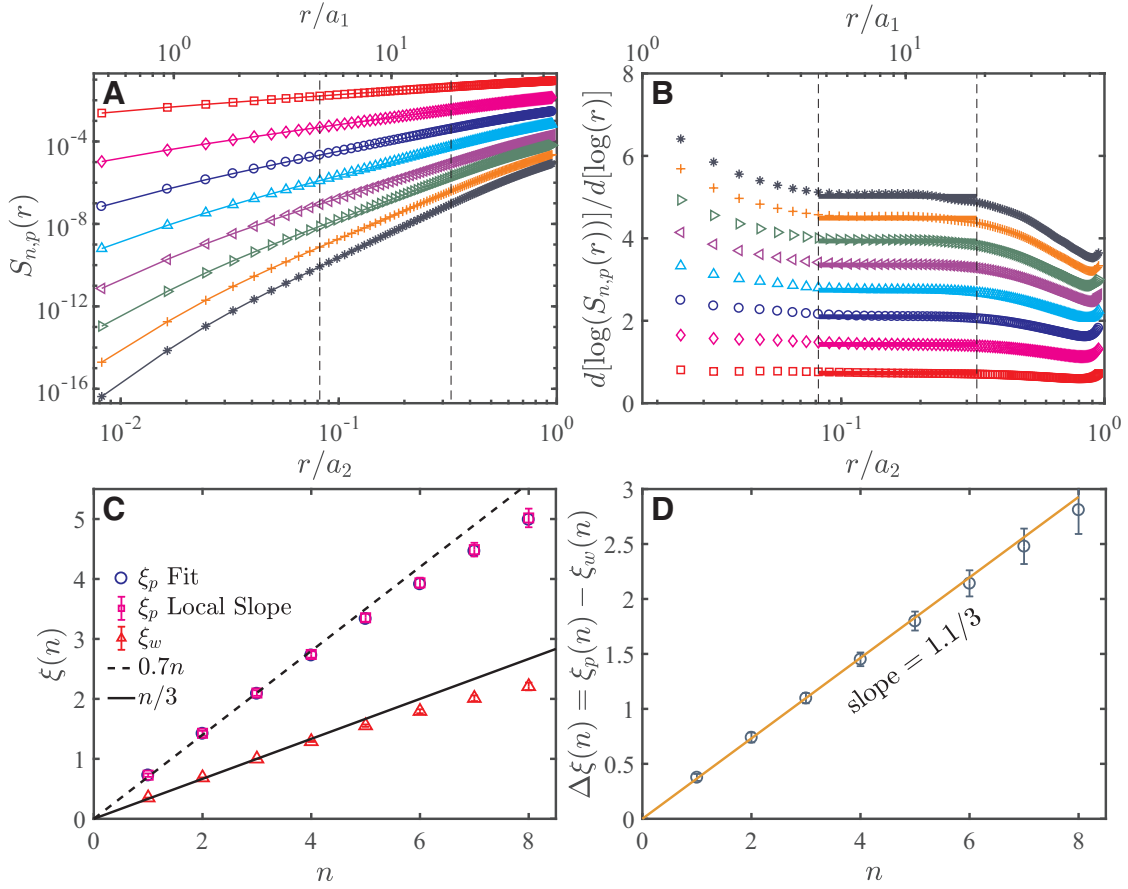}
	\caption{\textbf{Scaling of high order velocity structure function in the elastic range at $R_\lambda=480$ and $\phi = 40$ ppm.} (\textbf{A}) The \textit{n}th order ( $n = $1 to 8, from top to bottom) longitudinal VSF in the polymer solution $S_{n,p}(r)$ as a function of $r/a_2$ (or $r/a_1$, upper axis), the range between the two vertical dashed lines is the elastic range, the scaling exponent $\xi _p (n)$ is obtained from the power-law fitting to this range. The absolute values of the velocity increments are used to calculate the VSF. (\textbf{B}) Local slope ${d [\log (S_{n,p}(r))]}/{d [\log(r)]}$ of $S_{n,p}(r)$ for $n = $ 1 to 8 (from bottom to top) as a function of $r/a_2$ (or $r/a_1$, upper axis). The two vertical dashed lines mark the region where the local slope is nearly constant. The horizontal solid lines represent the average value within the two dashed lines. (\textbf{C}) Elastic range scaling exponents $\xi_p$ as a function of $n$. $\xi_p$ obtained from both the direct fitting and the local slope are plotted. The inertial range scaling exponents for pure water $\xi_w (n)$ is also plotted for comparison. The dashed line is  $\xi_p (n) = 0.7n$. The solid line is the K41 prediction, \textit{i.e.}, $\xi_w (n) = n/3$. (\textbf{D}) $\Delta \xi (n) = \xi _p (n)- \xi_w (n)$ as a function of $n$. The solid line is $\Delta \xi (n) =1.1n/3$.} \label{fig:high_order}
\end{figure}

\end{document}